\documentstyle[12pt,epsfig]{article}
\def\beq{\begin{equation}}
\def\eeq#1{\label{#1}\end{equation}}
\def\eeqn{\end{equation}}
\def\beqa{\begin{eqnarray}}
\def\eeqa#1{\label{#1}\end{eqnarray}}
\def\eeqan{\end{eqnarray}}
\def\CR{\nonumber \\ }
\def\leqn#1{(\ref{#1})}
\def\({\left(}
\def\){\right)}

\def\stacksymbols #1#2#3#4{\def\theguybelow{#2}
    \def\vp{\lower#3pt}
    \def\sp{\baselineskip0pt\lineskip#4pt}
    \mathrel{\mathpalette\intermediary#1}}

\def\intermediary#1#2{\vp\vbox{\sp
     \everycr={}\tabskip0pt
     \halign{$\mathsurround0pt#1\hfil##\hfil$\crcr#2\crcr
              \theguybelow\crcr}}}

\def\lapproxeq{\stacksymbols{<}{\sim}{2.5}{.2}}

\def\to{\rightarrow}
\def\eps{\epsilon}
\def\Phif{\Phi_{\rm FSR}}

\begin{document}

\begin{titlepage}
\begin{flushright}
{\tt hep-ph/0507194} \\
{\tt UFIFT-HEP-05-11}\\
\end{flushright}

\vskip.5cm
\begin{center}
{\huge \bf Robust Gamma Ray Signature of \\ \vskip.3cm WIMP Dark Matter} \vskip.2cm
\end{center}

\vskip1cm

\begin{center}
{\bf Andreas Birkedal$^{1}$, Konstantin T.~Matchev$^{1}$, Maxim Perelstein$^2$, \\
and Andrew Spray$^{2}$} \\
\end{center}
\vskip 8pt

\begin{center}
{\it $^1$ Physics Department, University of Florida, Gainesville, 
FL~32611\\
$^2$ Institute for High-Energy Phenomenology, Cornell University, 
Ithaca, NY~14853}
\end{center}

\vglue 0.3truecm

\begin{abstract}
\vskip 3pt \noindent
If dark matter consists of weakly interacting massive particles (WIMPs), annihilation of WIMPs in the galactic center may lead to an observable enhancement of high energy gamma ray fluxes. We predict the shape and normalization of the component of the flux due to final state radiation by charged particles produced in WIMP annihilation events. The prediction is made without any assumptions about the microscopic theory responsible for WIMPs, and depends only mildly on the unknown distribution of the total WIMP annihilation cross section among the possible final states. In particular, if the WIMPs annihilate into a pair of charged fermions (leptons or quarks), the photon spectrum possesses a sharp edge feature, dropping abruptly at a photon energy equal to the WIMP mass. 
%The shape of the edge can be predicted using collinear factorization %methods, and is completely model-independent. 
If such a feature is observed, it would provide strong evidence for the WIMP-related nature of the flux enhancement, as well as a measurement of the WIMP mass. We discuss the prospects for observational discovery of this feature at ground-based and space-based gamma ray telescopes.

\end{abstract}

\end{titlepage}

\section{Introduction} 

While the existence of dark matter on galactic and cosmological scales has 
been firmly established, its microscopic nature is still unknown. According to 
the ``WIMP hypothesis'', dark matter consists of stable, weakly interacting 
massive particles (WIMPs) with masses
roughly within the 10 GeV -- 10 TeV range. From the theoretical point
of view, this hypothesis is perhaps the most attractive among the
proposed candidate theories. There is as yet no direct evidence for its
validity; however, it does predict several potentially observable new
phenomena. In particular, pairs of WIMPs accumulated in the Milky Way and 
other galaxies should occasionally annihilate into lighter particles. These 
lighter particles (or their decay products) can then be found in cosmic rays, 
providing an ``indirect'' signature of galactic WIMPs. 

\begin{figure}[tb]
\begin{center}
\includegraphics[width=9.5cm]{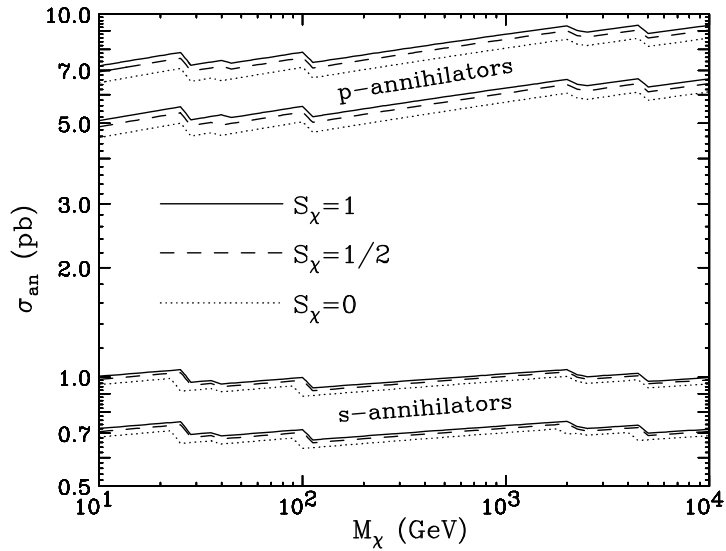}
\vskip2mm
\caption{Values of the quantity $\sigma_{\rm an}$ allowed at 2$\sigma$
level as a function of WIMP mass. The lower and upper bands correspond to models where the WIMP is an $s$- and $p$-annihilator, respectively. Reproduced from Ref.~\cite{us}.}
\label{fig:sigma}
\end{center}
\end{figure}

The same process, pair annihilation of WIMPs into lighter particles, is also responsible for maintaining the thermal equilibrium between the WIMPs and the rest of the cosmic fluid in the early universe. As a result, the temperature at which the WIMPs decouple depends sensitively on the pair annihilation cross section. This implies that
a measurement of the present dark matter density (currently known with an accuracy of about 10\%~\cite{wmap}) provides a determination of the  annihilation cross section under the conditions prevailing at the time of decoupling. Since WIMPs are non-relativistic at decoupling, it is useful to expand the total annihilation cross section as a power series in terms of the WIMP relative velocity $v$:
\beq
\sigma v = a \,+\, bv^2 \,+\,\ldots\,.
\eeq{nrexp}
In a generic situation, one of the two terms in this equation dominates the 
cross section at decoupling ($v^2\sim 3T/M \sim 0.1$): if $s$ wave 
annihilation is unsuppressed, the cross section is dominated by the $a$ term, 
whereas if the annihilation predominantly occurs in a $p$ wave, the $b$ term 
dominates. Therefore, a measurement of the present dark matter density 
determines the quantity $\sigma_{\rm an}$ defined in Ref.~\cite{us} as the 
coefficient of the dominant term (i.e. $\sigma_{\rm an}=a$ for 
$s$-annihilators and $\sigma_{\rm an}=b$ for $p$-annihilators). This result, 
shown in Fig.~\ref{fig:sigma}, is completely independent of the particle 
physics model responsible for the WIMPs; the only requirement is that the 
spectrum be generic, which ensures  that co-annihilation processes and 
resonances are unimportant\footnote{The analysis can also be extended to 
the case of superWIMP dark matter~\cite{Feng}.}. Moreover, 
$\sigma_{\rm an}$ is largely independent 
of the WIMP mass and spin: roughly, $\sigma^s_{\rm an}=0.85$ pb for 
$s$-annihilators and $\sigma^p_{\rm an}=7$ pb for $p$-annihilators.

In this article, we extend the model-independent approach of Ref.~\cite{us} 
to predict the fluxes of anomalous cosmic rays due to WIMP annihilation. 
Indirect WIMP searches predominantly concentrate on three signatures: 
anomalous high-energy gamma rays, antimatter (positrons, antiprotons, etc.), 
and neutrinos~\cite{review}. While the dark matter density measurement 
determines the {\it total} cross section of WIMP annihilation, the 
distribution between the various possible final states ($e^+e^-$, 
$q\bar{q}$, $\gamma\gamma$, $W^+W^-$, etc.) is not constrained. In order to 
keep the analysis as model-independent as possible, we focus on the 
signatures that are least sensitive to this distribution, i.e. those that 
appear for the maximal number of final states. High-energy neutrinos and 
positrons are only produced if the WIMPs annihilate directly into 
$\nu\bar{\nu}$ or $e^+e^-$ pairs, respectively, or (in smaller numbers) if 
the primary annihilation final state contains $W/Z$ bosons. Gamma rays, on 
the other hand, are produced almost independently of the primary final state 
(with $\nu\bar{\nu}$ being the only exception among two-body final states), 
and we therefore concentrate on this signature. There are several ways in 
which gamma rays can be produced in WIMP annihilation events. Two well known 
processes~\cite{previous,SUSY2gamma,Bergs,UEDdm} are the direct annihilation 
to photon-photon or photon-$Z$ pairs ($\chi\chi\to \gamma\gamma, \gamma Z$, 
where $\chi$ denotes the WIMP) and fragmentation following WIMP annihilation 
into final states containing quarks and/or gluons. While these processes can 
be easily described within our approach, we will concentrate on another source 
of photons, the final state radiation (FSR), which has until now received 
far less attention in the literature\footnote{A recent discussion of the FSR 
flux in the context of a specific model (universal extra dimensions) and a 
subset of primary final states (charged leptons) is contained 
in~\cite{BergsUED}.}. The FSR component of the gamma ray spectrum has several 
important advantages. First, FSR photons are produced whenever the primary 
products of WIMP annihilation are charged: e.g. charged leptons, quarks or 
$W$ bosons. Even if the WIMPs annihilate into $ZZ$, $Zh$, or $hh$ pairs, the 
charged decay products of these particles will contribute to the FSR flux; 
only the $\nu\bar{\nu}$ channel does not contribute. In contrast, the
monochromatic photons are only produced when the WIMPs annihilate into
$\gamma\gamma$ or $\gamma Z$ pairs; since these processes can only occur at 
one-loop level~\cite{SUSY2gamma}, only a small fraction of WIMP annihilation 
events results in these funal states. The fragmentation photons are not 
produced for leptonic final states. In this sense, out of the three components 
of the photon flux, the FSR component is the most robust. 
Second, even though the energy spectrum of the FSR photons is broad, in many 
cases (whenever the WIMPs annihilate directly into charged fermion pairs) the 
spectrum contains a sharp edge feature at an energy close to the WIMP 
mass~\cite{BergsUED}. This feature can be extremely useful in differentiating 
the WIMP signal from the astrophysical background: while no detailed 
theoretical understanding of the background is available, it seems very 
unlikely that such a feature in the relevant energy range can be produced by 
conventional physics. This is in sharp contrast with the fragmentation 
photons, whose broad and featureless spectrum makes it difficult to rule out 
a more conventional astrophysical explanation if an excess over the expected 
background is observed.

This article is organized as follows. In Section~\ref{FSR} we present 
the model-independent approximate formulas for the energy spectrum of the 
FSR photons produced in WIMP annihilation events. We test the accuracy of 
our analytical results against explicit numerical calculations in specific 
models. In Section~\ref{sec:flux}, we use these results to predict the gamma 
ray fluxes from WIMP dark matter annihilation in the Milky Way. After 
discussing the relevant backgrounds in Section~\ref{bg}, we estimate the 
sensitivity reach of the typical space-based and ground-based gamma ray 
telescopes in Section~\ref{reach}. We reserve Section~\ref{conc} for our 
conclusions.

\section{Final State Radiation in WIMP Annihilation}
\label{FSR}

If a WIMP pair can annihilate into a pair of charged particles, $X$ and 
$\bar{X}$, annihilation into a three-body final state $X\bar{X}\gamma$ is 
always also possible. As long as the $X$ particles in the final state are 
relativistic, the cross section of this reaction is dominated by the photons 
that are approximately {\it collinear} with either $X$ or $\bar{X}$. These 
are referred to as the ``final state radiation'' (FSR) photons. In this 
kinematic regime, the cross section factorizes into the short-distance part, 
$\sigma(\chi\chi\to X\bar{X})$, and a universal collinear factor: 
\beq
\frac{d\sigma (\chi\chi\,\to\,X\bar{X}\gamma)}{dx} \approx  
\frac{\alpha Q_X^2}{\pi}\,{\cal F}_X(x)\,\log\left(\frac{s(1-x)}{m_X^2}\right) 
\sigma (\chi\chi\to X\bar{X}),
\eeq{wwapprox}
where $\alpha$ is the fine structure constant, $Q_X$ and $m_X$ are the 
electric charge and the mass of the $X$ particle, $s$ is the center-of-mass 
energy ($s\approx 4m_\chi^2$ for non-relativistic WIMPs), and 
$x=2E_\gamma/\sqrt{s}$. The splitting function ${\cal F}$ is independent of 
the short-distance physics, depending only on the spin of the $X$ particles. 
If $X$ is a fermion, the splitting function is given by
\beq
{\cal F}_f(x) \,=\, \frac{1 + (1-x)^2}{x}, 
\eeq{splitf}
whereas if $X$ is a scalar particle,
\beq
{\cal F}_s(x) \,=\, \frac{1-x}{x}.
\eeq{splits}
If $X$ is a $W$ boson, the Goldstone boson equivalence theorem implies that 
${\cal F}_W(x)\approx {\cal F}_s(x)$. (The applicability of the Goldstone 
boson equivalence theorem is guaranteed whenever the collinear factorization 
in Eq.~\leqn{wwapprox} is a good approximation, since both require 
$m_\chi\gg m_W$.) 

\begin{figure}[tb]
\begin{center}
\includegraphics[width=9.5cm]{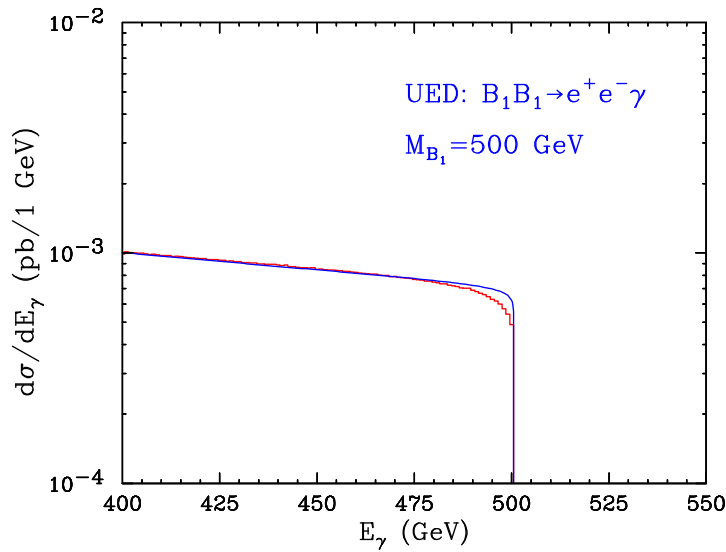}
\vskip2mm
\caption{Comparison of the photon spectrum obtained by a direct 
calculation in the UED model with the radius of the extra dimension $R=(499.07$ GeV$)^{-1}$ (red histogram) and the spectrum predicted by Eq.~\leqn{wwapprox} (blue line) for the case of $B_1B_1\to e^+e^-\gamma$ annihilation at $\sqrt{s}=1001$ GeV. The mass of the lightest Kaluza-Klein particle (the first excited mode $B_1$ of the hypercharge gauge boson) is 500 GeV.}
\label{fig:ee}
\end{center}
\end{figure}

%The above equations predict the spectrum of the FSR photons emerging from the WIMP annihilation event. At the relevant energies, the photons propagate through the galactic interstellar gas without interacting, so the spectrum observed by a gamma ray telescope would be exactly the same. 

Does Eq.~\leqn{wwapprox} provide a good approximation of the FSR photon 
spectrum from galactic WIMP annihilation in a realistic situation? To address 
this question, we compare the FSR photon spectrum obtained by a direct 
calculation in a specific model with the prediction of Eq.~\leqn{wwapprox} 
with the appropriate parameters. For this comparison, we have used the 
minimal universal extra dimension (UED) model~\cite{UED}. We computed the 
cross section of the process $B_1B_1\to e^+e^- \gamma$ using the {\tt CompHEP} 
package~\cite{comphep}. ($B_1$, the first Kaluza-Klein excitation of the 
hypercharge gauge boson, plays the role of the WIMP dark matter candidate in 
the UED model~\cite{KKDM}.) We have fixed the radius of the extra dimension to 
be $R=(499.07$ GeV$)^{-1}$, corresponding to $B_1$ mass of 500 GeV. While 
Eq.~\leqn{wwapprox} holds for any WIMP momentum, we have chosen the colliding
WIMPs to be nonrelativistic ($\sqrt{s}=1001$ GeV), to approximate the 
kinematics typical of galactic WIMP collisions. The result of the direct 
cross section calculation is shown by the red histogram in Fig.~\ref{fig:ee}. 
The blue (continuous) line corresponds to the prediction of 
Eq.~\leqn{wwapprox} with the same $\sqrt{s}=1001$ GeV, $X=e$, and the 
appropriate value of $\sigma(\chi\chi\to e^+e^-)\approx 5.67$ pb. 
The good agreement between the line and the histogram proves the validity of 
the collinear approximation for the total cross section.
Remarkably, the spectrum has a sharp step-like edge feature at the 
endpoint, $E\to M_\chi$. The origin of the feature is obvious from 
Eqs.~\leqn{wwapprox} and~\leqn{splitf}: ignoring the $x$ dependence of the 
logarithm in Eq.~\leqn{wwapprox}, which only has a small effect on the 
spectrum, it is easy to see that the differential cross section approaches a 
non-zero constant value at $x\to 1$, whereas it obviously has to vanish for 
$x>1$. Since it is difficult to imagine an astrophysical process providing a 
similarly sharp endpoint feature at the relevant energy scales, observing the 
step would provide a strong evidence for WIMPs~\cite{BergsUED}.

\begin{figure}[tb]
\begin{center}
\includegraphics[scale=0.47]{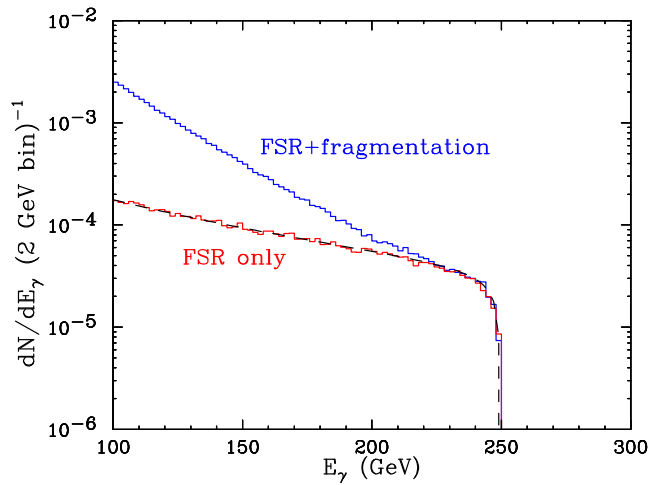}
\vskip2mm
\caption{Photon spectrum produced by final state radiation and fragmentation of a primary $u$ quark with an energy of 250 GeV. 
The histograms represent {\tt PYTHIA} predictions for the total photon flux (blue) and the final state radiation flux alone (red). The black dashed line represents the prediction of Eq.~\leqn{wwapprox}.}
\label{fig:uu}
\end{center}
\end{figure}

If the primary product of WIMP annihilation is a lepton pair ($e^+e^-$, 
$\mu^+\mu^-$), the FSR mechanism discussed above is the dominant source of 
secondary photons. On the other hand, if the WIMPs annihilate into 
quark-antiquark or $\tau^+\tau^-$ pairs, an additional contribution to the 
secondary photon flux arises from hadronization and fragmentation. This 
contribution is dominated by the decays of neutral pions. While the 
fragmentation photons are more numerous than the FSR photons, they tend to be 
softer. The spectrum close to the endpoint is still dominated by the FSR 
component, and can be predicted using Eq.~\leqn{wwapprox}, with an appropriate 
choice of the ``effective'' value of $m_X$ in the logarithm. This is 
illustrated in Figure 3, which shows the secondary photon fluxes from a 
primary $u$ quark of 250 GeV energy. The upper (blue) histogram shows the 
total $\gamma$ spectrum, including both fragmentation and FSR components, 
calculated using the {\tt PYTHIA} package~\cite{pythia}; the lower (red) 
histogram shows the 
{\tt PYTHIA} prediction for the FSR flux alone. The total flux above about 
225 GeV ($x=0.9$) is dominated by the FSR component, and exhibits the expected 
edge feature. The FSR spectrum predicted by {\tt PYTHIA} is consistent with 
the prediction of Eq.~\leqn{wwapprox}; however, to obtain a good fit, the 
quantity $m_u$ in the logarithm should be replaced with the ``effective mass''
$m_u^{\rm eff}$, which takes into account soft gluon radiation and other 
effects of strong interactions.  The black dashed line in Fig.~\ref{fig:uu} 
represents the prediction of Eq.~\leqn{wwapprox} using the best-fit value of 
$m_u^{\rm eff}=20$ GeV. An excellent fit to the {\tt PYTHIA} output is 
obtained. We conclude that the sharp endpoint with a shape given by 
Eqs.~\leqn{wwapprox},~\leqn{splitf} exists whenever the primary WIMP 
annihilation products are charged fermions: it does not matter whether they 
are leptons or quarks.

Based on the above discussion, we will replace the bare mass with the 
``effective'' mass whenever we apply Eq.~\leqn{wwapprox} to light quarks; 
this substitution will be implicit for the rest of the paper. For simplicity, 
we will assume $m_u^{\rm eff}=m_d^{\rm eff}=m_s^{\rm eff}=0.2$ GeV, since the 
scale for the effective mass is set by the QCD confinement scale. In reality, 
the situation is more complicated, since the effective mass may depend on the 
energy and flavor of the primary quark. Note also that the values needed to 
fit the {\tt PYTHIA} predictions are substantially higher than 
$\Lambda_{\rm QCD}$, but the interpretation of this result is not clear due
 to large uncertainties inherent in the showering algorithm and the fit. 
However, since the dependence on the mass is logarithmic, changing 
$m^{\rm eff}$ has only a moderate effect on the photon flux predictions. We 
have confirmed that replacing our simple assumption with an energy-dependent 
value of $m^{\rm eff}$ based on a fit to {\tt PYTHIA} predictions does not 
substantially affect any of the estimates of photon fluxes and telescope 
sensitivities made below.

Unfortunately, a model-independent prediction of the sharp endpoint is not 
valid if the primary annihilation products of the WIMP are bosons, such as 
$W^+W^-$ pairs or the charged Higgs bosons in the minimal supersymmetric 
standard model (MSSM). According to Eq.~\leqn{splits}, 
$\lim_{x\to1}{\cal F}_s(x)=0$; because of this, the flux near the endpoint is 
dominated by the model-dependent non-collinear contributions, and no firm 
model-independent prediction of the shape of the endpoint spectrum is possible.

FSR photons will also be produced when the WIMPs annihilate into neutral, 
unstable particles, whose decay products are charged: in the Standard Model, 
these could be $ZZ$, $Zh$, or $hh$ pairs. 
%Annihilation into such final states is likely in SUSY theories with non-%universal gaugino masses\cite{NonUniversal}.  
For example, consider the process $\chi\chi\to ZZ$ with the subsequent $Z$ 
decay into charged fermions ($\ell^+\ell^-$ or $q\bar{q}$), which in turn 
emit an FSR photon. The photon spectrum in the $Z$ rest frame is given by 
Eq.~\leqn{wwapprox}, with the substitutions $s\to m_Z^2$ and 
$\sigma(\chi\chi\to X\bar{X}) \to 2\sigma(\chi\chi\to ZZ) {\rm Br}
(Z\to X\bar{X})$. Performing the boost to return to the laboratory frame 
yields
\beq
\frac{d\sigma}{dx} \,=\, \frac{\alpha}{\pi}\,\sigma(\chi\chi\to ZZ) \,
\Psi_Z(x),
\eeq{4body}
where
\beq
\Psi_Z = 2 \theta(m_\chi-m_Z)
\,\frac{1}{x}\,\left[ 1+\frac{1-x}{v} - \frac{2x^2}{v(v+1)}
+ \frac{2x}{v}\log \frac{2xv}{1+v} \right] \sum_X Q_X^2\,
{\rm Br}(Z\to X\bar{X})\,
\log\left(\frac{m_Z^2}{m_X^2}\right).
\eeq{PsiZexact}
In this equation, $x=E_\gamma/m_\chi$, $v=\sqrt{1-m_Z^2/m_\chi^2}$ is the 
velocity of the $Z$ boson, 
and the sum runs over all the charged fermion pairs that $Z$ can decay into.
We have ignored the corrections that are not enhanced by 
$\log(m_Z^2/m_X^2)$. If $m_\chi\gg m_Z$, the $Z$ bosons are relativistic and
the spectrum is given by
\beq
\Psi_Z = 2 
\,\frac{2-x+2x\log x-x^2}{x}\,\sum_X Q_X^2\,{\rm Br}(Z\to X\bar{X})\,
\log\left(\frac{m_Z^2}{m_X^2}\right).
\eeq{PsiZ}
For the $Zh$ and $hh$ final states, we obtain expressions analogous 
to~\leqn{4body}. The corresponding functions $\Psi_{Zh}$ and $\Psi_h$ can be 
easily obtained from Eq.~\leqn{PsiZexact} by replacing the parameters of the 
$Z$ boson with those of the Higgs where appropriate. In particular, in the 
limit $m_\chi\gg m_h$ we obtain
\beqa
\Psi_{hZ}(x) &=& \frac{2-x+2x\log x - x^2}{x}\,\sum_X Q_X^2 \Bigl[ {\rm Br}(Z\to X\bar{X})\, \log\left(\frac{m_Z^2}{m_X^2}\right) \CR & & \,+
{\rm Br}(h\to X\bar{X}) \, \log\left(\frac{m_h^2}{m_X^2}\right)
\Bigr]\,, \CR
\Psi_h(x) &=& 2\, \frac{2-x+2x\log x - x^2}{x}\,\sum_X Q_X^2 \,{\rm Br}(h\to X\bar{X}) \, \log\left(\frac{m_h^2}{m_X^2}\right)\,.
\eeqa{Psifunctions}
These expressions include only fermionic decays of the Higgs; we assumed that 
the Higgs is too light to decay into $W/Z$ pairs. The analysis can be 
straightforwardly generalized to include these decays. Unfortunately, it is 
clear from the above equations that the spectrum of FSR photons produced in 
$Z/h$ decays does not possess a sharp endpoint; instead, it approaches 0 
gradually in the $x\to1$ limit. This means that the non-universal, 
model-dependent contributions may become dominant near the endpoint. 

\section{FSR Photon Flux Estimates} 
\label{sec:flux}

In general, the differential $\gamma$ flux from WIMP annihilations observed 
by a telescope can generally be written as
\beq
\frac{d^2\Phi}{dE d\Omega} = \left( \sum_{i}  \left< \frac{d\sigma_i}{dE} v \right> \,B_i \right)\, \frac{1}{4\pi m^2_\chi} \int_\Psi \rho^2(l) dl,
\eeq{flux0}
where the sum runs over all possible annihilation channels containing photons, 
and $\sigma_i$
 and $B_i$ are the annihilation cross section and the number 
of photons per event in a given channel, respectively. The average is over 
the thermal ensemble of WIMPs in the galaxy. The integral is 
computed along a line of sight in the direction parametrized by 
$\Psi = (\theta, \phi)$, and $\rho(l)$ is the mass density of WIMP dark 
matter at a distance $l$ from the observer. To obtain the FSR photon flux, we 
substitute the differential cross section for the 
%$X\bar{X}\gamma$ 
final states containing such photons, given in Eqs.~\leqn{wwapprox} 
and~\leqn{4body}, into Eq.~\leqn{flux0}, and take into account that $B_i=1$ 
for these final states. We obtain 
\beqa
\frac{d^2\Phif}{dEd\Omega} &=&  \Bigl[\sum_X \theta(m_\chi-m_X) Q_X^2 \left< \sigma_X v\right>
{\cal F}_X(x) \log\left(\frac{4m_\chi^2(1-x)}{m_X^2}\right)\,+\, \left<\sigma_Z v\right> \Psi_Z(x) \CR & &\,+\, \left<\sigma_{hZ} v\right> \Psi_{hZ}(x)\,+\,\left<\sigma_h v\right> \Psi_h(x) \Bigr]\,\,
\times \frac{\alpha}{\pi} \frac{1}{4\pi m_\chi^3} \int_\Psi \rho^2 dl,
\eeqa{fluxFSR}
where $x=E/m_\chi$. The sum runs over all possible two-body final states with 
charged particles $X$ and $\bar{X}$, and 
$\sigma_X=\sigma(\chi\chi\to X\bar{X})$. To simplify notation, we have also 
defined $\sigma_Z=\sigma(\chi\chi\to ZZ)$, $\sigma_h=\sigma(\chi\chi\to hh)$, 
and $\sigma_{hZ}=\sigma(\chi\chi\to Zh)$. 

In the spirit of Ref.~\cite{us}, we define the total WIMP annihilation cross 
section, $\sigma_0 = \sigma(\chi\chi\to {\rm anything})$, and the 
``annihilation fractions'' for the two-particle final states, 
$\kappa_X=\left< \sigma_X v\right>/ \left<\sigma_0 v\right>$. (Note that 
$\sum_X \kappa_X = 1$, up to a small correction due to the contribution of 
the processes with three or more particles in the final state.) With these 
definitions, the FSR flux can be written as 
\beq
\frac{d^2\Phif}{dEd\Omega}\,=\,\frac{\alpha}{\pi} \frac{1}{4\pi m_\chi^3} 
\left< \sigma_0 v\right> {\cal G}(x) \int_\Psi \rho^2 dl,
\eeq{fluxFSR1}
where
\beq
{\cal G}(x)= \sum_X \theta(m_\chi-m_X) Q_X^2 \kappa_X {\cal F}_X(x) 
 \log\left(\frac{4m_\chi^2(1-x)}{m_X^2}\right)\,+\,\kappa_Z\Psi_Z(x)\,+\, 
 \kappa_{hZ} \Psi_{hZ}(x)\,+\,\kappa_h \Psi_h(x).
\eeq{Gdef}
Notice that almost all WIMP annihilation channels, with the exception of 
$\nu\bar{\nu}$ and $gg$ final states, contribute to the FSR photon flux; only 
the details of the flux depend on the distribution of the cross section among 
the channels. 

The photon flux prediction is subject to large uncertainties in the 
distribution of dark matter in the galaxy. These uncertainties are 
conventionally parametrized by a dimensionless function
\beq
\bar{J}(\Psi, \Delta\Omega ) \equiv \frac{1}{8.5~{\rm kpc}} \left( 
\frac{1}{0.3 {\rm~GeV/cm}^3} \right)^2 \frac{1}{\Delta\Omega} 
\int_{\Delta\Omega} d\Omega \int_\Psi \rho^2 dl,
\eeq{3}
where $\Delta\Omega$ denotes the field of view of a given experiment. The 
values of $\bar{J}$ depend on the galactic halo model. The optimal line of 
sight for WIMP searches is towards the galactic center; in this case, the 
uncertainty is particularly severe due to the possibility of a sharp density 
enhancement at the center. For example, at $\Delta\Omega=10^{-3}$ sr, typical 
values of $\bar{J}$ range from $10^3$ for the NFW profile~\cite{NFW} to 
about $10^5$ for the profile of Moore et.al.~\cite{Moore}, and can be further 
enhanced by a factor of up to $10^2$ due to the effects of adiabatic 
compression~\cite{abc}.

Using Eq.~\leqn{fluxFSR1} and the above definition of $\bar{J}$ yields the 
FSR flux integrated over the field of 
view: 
\beq
\frac{d\Phif}{dE} =  \Phi_0 \left(\frac{\left<\sigma_0 v\right>}{1~{\rm pb}}
\right)\,\left(\frac{100~{\rm GeV}}{m_\chi}\right)^3\, {\cal G}(x)
\bar{J}(\Psi, \Delta\Omega) \Delta\Omega,
\eeq{flux}
where $\Phi_0=1.4\times 10^{-14}$ cm$^{-2}$ sec$^{-1}$ GeV$^{-1}$.

While Eq.~\leqn{flux} provides a complete description of the FSR photon 
spectrum, the shape and the normalization of the flux for the most energetic 
photons (close to $x=1$) is of particular interest due to the possibility of 
the observable edge feature. In this region, the flux is dominated by the 
photons radiated by fermion products of WIMP annihilation. Neglecting the $x$ 
dependence in the logarithm, whose only effect is to slightly smooth out the 
edge in the region $1-x\ll1$, the flux is approximately given by
\beq
\frac{d\Phif}{dE} =  \Phi_0\,g\,
\left(\frac{100~{\rm GeV}}{m_\chi}\right)^3\, {\cal F}_f(x)
\bar{J}(\Psi, \Delta\Omega) \Delta\Omega.
\eeq{flux_edge}
Here, the dimensionless parameter $g$ contains all the information about the 
primary WIMP annihilation processes:
\beq
g \,=\, \left( \frac{\left<\sigma_0 v\right>}{1~{\rm pb}}\right)\,
\sum_f Q_f^2
\kappa_f \log\left(\frac{4m_\chi^2}{m_f^2}\right),
\eeq{gdef}
where the sum runs over all kinematically accessible fermionic final states. 
Depending on the microscopic model giving rise to the WIMP, the parameter $g$ 
can vary between 0 (if, for example, the WIMPs can only annihilate into 
neutral states) and about 35 in the most favorable case of very heavy WIMPs 
annihilating into electron-positron pairs in an $s$ wave. 

The normalization of the FSR photon flux is determined by the quantity 
$\left<\sigma_0 v\right>$. As we argued in the Introduction, the measurement 
of the present cosmological abundance of dark matter determines the total WIMP 
annihilation cross section at decoupling ($v^2\sim1/10$). A typical relative 
velocity of galactic WIMPs is much smaller, $v\sim 10^{-3}$. In models where 
the $s$-wave annihilation is unsuppressed, the quantity $\sigma v$ is 
velocity-independent at low $v$, allowing us to make a robust 
model-independent prediction:
\beq
\left<\sigma_0 v\right> \,=\,\sigma^{s}_{\rm an} \,\approx\,0.85{~\rm pb}.
\eeq{sigma0v}
If, on the other hand, the cross section at decoupling is dominated by the $b$ 
term, no firm prediction for the quantity $\left<\sigma_0 v\right>$ is 
possible: even a small $a$ term, if present, may  become dominant for galactic 
WIMPs due to the low value of $v$. If no $a$ term is present, we estimate 
$\left<\sigma_0 v\right> = \sigma^p_{\rm an} v^2 \sim 10^{-5}$ pb; a larger 
cross section is possible if an $a$ term is present, with the upper bound 
provided by Eq.~\leqn{sigma0v}. Given the uncertainty present in the 
$p$-annihilator case, we will use the $s$-annihilator WIMP examples to 
illustrate our approach in the remainder of this paper. 

\section{Background Fluxes}
\label{bg}

Estimating the sensitivity of WIMP searches also requires the knowledge of 
background fluxes. The searches for anomalous cosmic $\gamma$ rays are 
conducted both by space-based telescopes and ground-based atmospheric Cerenkov 
telescopes (ACTs). The space-based telescopes observe photons directly, and 
the only source of irreducible background in this case is the cosmic $\gamma$ 
rays of non-WIMP origin. The ACTs observe the Cerenkov showers created when a 
cosmic ray strikes the upper atmosphere, and are subject to the additional 
backgrounds of Cerenkov showers from leptonic and hadronic cosmic rays. 
%(Leptonic showers are identical to photonic showers, and can only %be distinguished based on direction; hadronic showers are different %in shape and can be distinguished to some accuracy.) 
In our estimates of the experiments' sensitivities, we will use simple 
power-law extrapolations of the background 
fluxes measured at low energies. For the non-WIMP photon flux, we 
use~\cite{Bergs} 
\beq
\frac {d^2\Phi_{\gamma,~{\rm bg}}}{dEd\Omega} = 4 \times 10^{-12} 
\,N_0(\Psi)~\left(\frac {100~{\rm GeV}}{E} \right)^{2.7}{\rm~cm}^{-2} 
{\rm s}^{-1} {\rm GeV}^{-1} {\rm sr}^{-1},
\eeq{sst_bgd}
where the function $N_0$ describes the angular distribution of the photons 
(an approximation is given in Refs.~\cite{Bergs,Nfun}.) In our analysis, we 
will replace $N_0(\Psi) \to \max N_0 \approx 89$. This generally overestimates 
the background; however, the effect is small, especially for the line of sight 
close to the direction to the galactic center. The non-photonic 
background flux for the ACTs is estimated as~\cite{Bergs}
\beqa
\frac {d^2\Phi_{\rm lep}}{dEd\Omega} &=& 1.73 \times 10^{-8} \left( \frac 
{100{\rm~GeV}}{E} \right)^{3.3} {\rm~cm}^{-2}{\rm s}^{-1}{\rm GeV}^{-1} 
{\rm sr}^{-1},\CR
\frac {d^2\Phi_{\rm had}}{dEd\Omega} &=&  4.13 \times 10^{-8} \,
\epsilon_{\rm had}\,\left( \frac {100{\rm~GeV}}{E} \right)^{2.7}{\rm~cm}^{-2}
{\rm s}^{-1}{\rm GeV}^{-1} {\rm sr}^{-1}, \CR
\eeqa{act_bgd}
where $\epsilon_{\rm had}$ is the telescope-dependent probability that a 
hadronic Cerenkov shower will be misidentified as a photonic shower, 
normalized so that it is equal to one for the Whipple telescope (see
Ref.~\cite{Bergs}).

It is worth emphasizing that the fluxes~\leqn{sst_bgd} 
and~\leqn{act_bgd} are merely extrapolations; in both cases the background 
cannot be accurately predicted from theory. While we will use these fluxes in 
our estimates, one should keep in mind that there are large uncertainties 
associated with them. This is why merely observing a flux enhancement 
is in general not sufficient to provide convincing evidence for WIMPs; a
discovery of the step-like edge feature in the spectrum would greatly 
strengthen the case.

\section{Sensitivity Reach of Future Telescopes}
\label{reach}

To illustrate the prospects for observational discovery of the FSR edge, we 
will use two toy scenarios. In the first scenario, the annihilation fractions 
for two-body final states are taken to scale as $Y^4N_c$, where $Y$ is the 
hypercharge of the final state particles, and $N_c=3$ for quarks and $1$ for 
other states\footnote{Note that Eq.~\leqn{wwapprox} can be applied to 
polarized final states. Therefore, accounting for the different hypercharge of 
left-handed and right-handed fermions is straightforward.}. An explicit 
example in which this scenario is realized is provided by the model with 
universal extra dimensions~\cite{UED}, and we will therefore label it as UED. 
In the second scenario, the WIMPs do not annihilate into bosonic final states, 
while the annihilation fractions $\kappa_i$ for all kinematically accessible 
fermion final states are equal (up to a factor of $N_c$). We will refer to 
this scenario as ``democratic''. In both scenarios, we assume that the WIMPs 
can only annihilate into Standard Model particles, and use a Higgs mass of 120 
GeV. The values of the quantity $g$, defined in Eq.~\leqn{gdef}, as a function 
of the WIMP mass $m_\chi$, in the two scenarios under consideration are shown 
in Figure~4. In both cases, we assume that WIMPs are 
$s$-annihilators, with the total annihilation cross section given by 
Eq.~\leqn{sigma0v}. 

\begin{figure}[tb]
\begin{center}
\includegraphics[width=9.5cm]{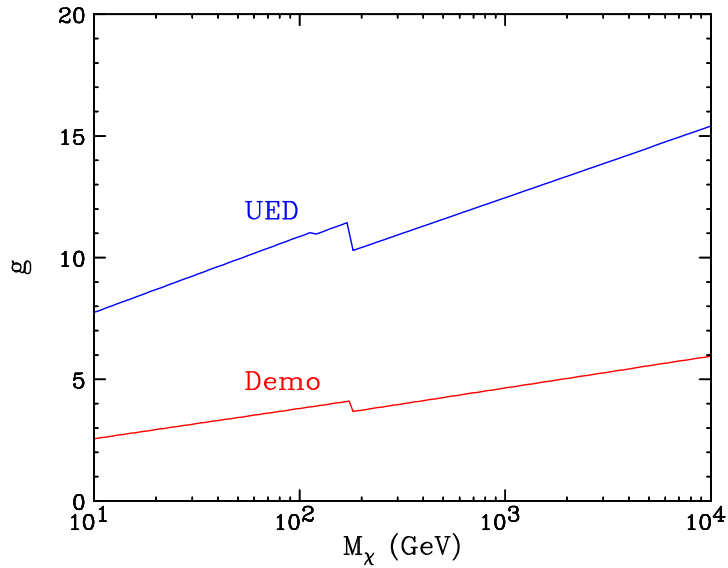}
\vskip2mm
\caption{The quantity $g$, defined in Eq.~\leqn{gdef}, as a function of 
the WIMP mass $m_\chi$, in the UED scenario (blue line) and the "democratic" 
scenario (red line). In the UED scenario, the annihilation fractions for 
two-body final states are taken to scale as $Y^4N_c$, where $Y$ is the 
hypercharge of the final state particles, and $N_c=3$ for quarks and $1$ 
for other states. In the second scenario, the annihilation fractions for 
all kinematically accessible two-fermion final states are equal (up to a 
factor of $N_c$).}
\end{center}
\label{fig:gg}
\end{figure} 

\begin{figure}[t]
\begin{center}
\includegraphics[width=9.5cm]{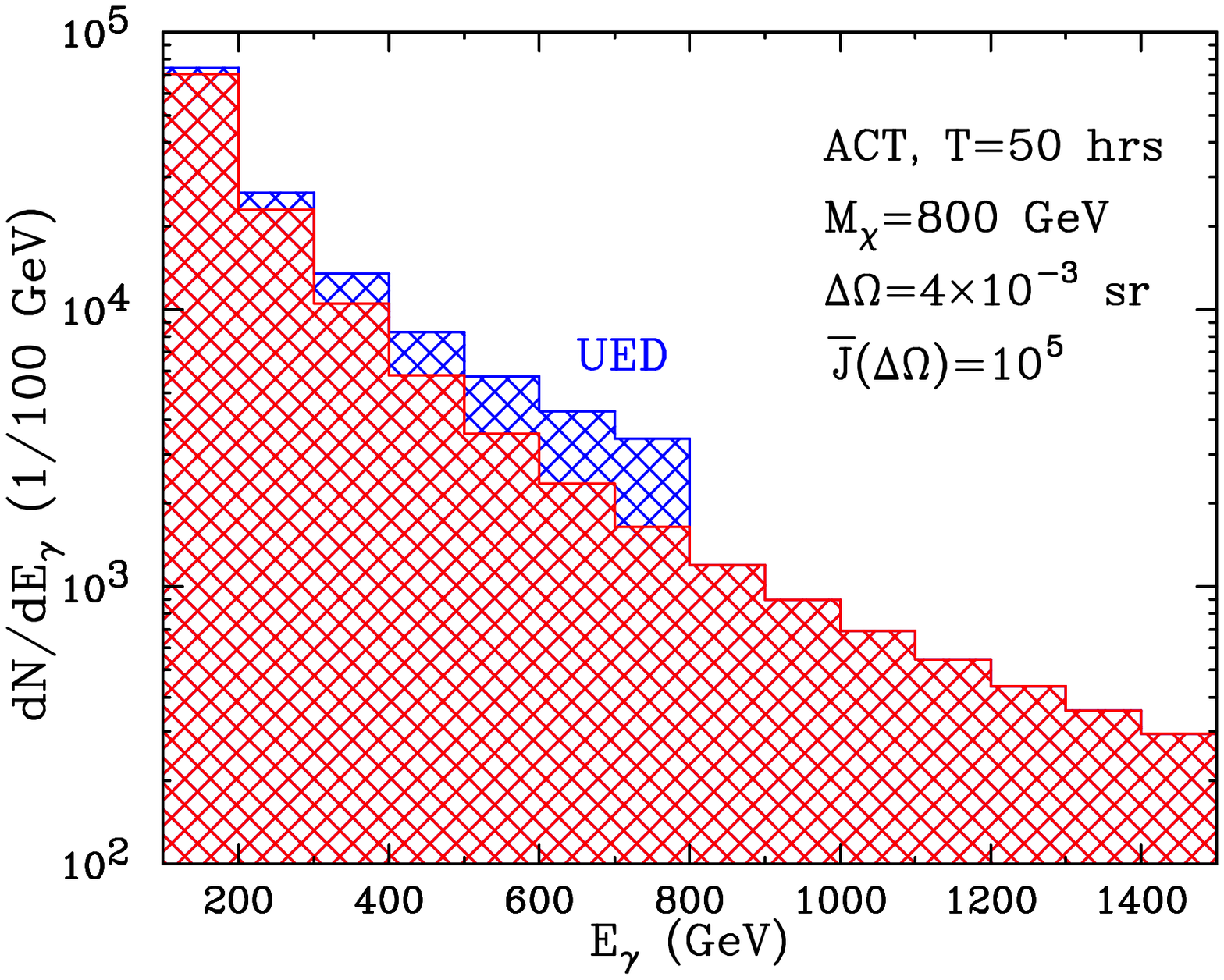}
\vskip2mm
\caption{The number of signal (blue) and background (red) events at a 
representative atmospheric Cerenkov telescope with a collection area given in
Eq.~\leqn{effarea}, an exposure time $T=50$ hrs, and a field of view 
$\Delta\Omega=4\times 10^{-3}$ sr. The signal is computed assuming the 
UED scenario with an 800 GeV WIMP and a galactic model with 
$\bar{J}(\Delta\Omega)=10^5$.}
\end{center}
\label{fig:ACT}
\end{figure} 

The magnitude of the FSR photon flux in each scenario is easily estimated 
using Eq.~\leqn{flux}. As an example, Fig.~5 shows the number of events per 
100 GeV bin expected to be observed at an ACT with an exposure time $T=50$ hrs,
and a field of view $\Delta\Omega=4\times 10^{-3}$ sr. (These parameters are 
similar to those of the VERITAS~\cite{veritas} and HESS~\cite{HESS} telescope 
arrays.) The effective collection area of the ACTs depends on the photon 
energy; in our analysis, we use an analytic fit to the effective area of the
VERITAS array shown in Fig.~4, Ref.~\cite{veritas}:
\beq
A(E)\,=\,1.2 \exp\left[-0.513\left(\log\frac{E}{5~{\rm TeV}}\right)^2 \right] 
\times 10^9 {~\rm cm}^2.
\eeq{effarea}
We assumed 
the UED scenario with an 800 GeV WIMP. We have further assumed 
$\bar{J}(\Delta\Omega)=10^5$, which is the case in the NFW galactic 
profile~\cite{NFW} with an adiabatic compression enhancement factor of about 
a 100~\cite{abc}, or in the profile of Moore et. al.~\cite{Moore} with no 
adiabatic compression. It is clear from the figure that the edge feature due 
to the FSR photon emission following WIMP annihilation should be easily 
discernible in this data set.

\begin{figure}[tb]
\begin{center}
\includegraphics[width=9.5cm]{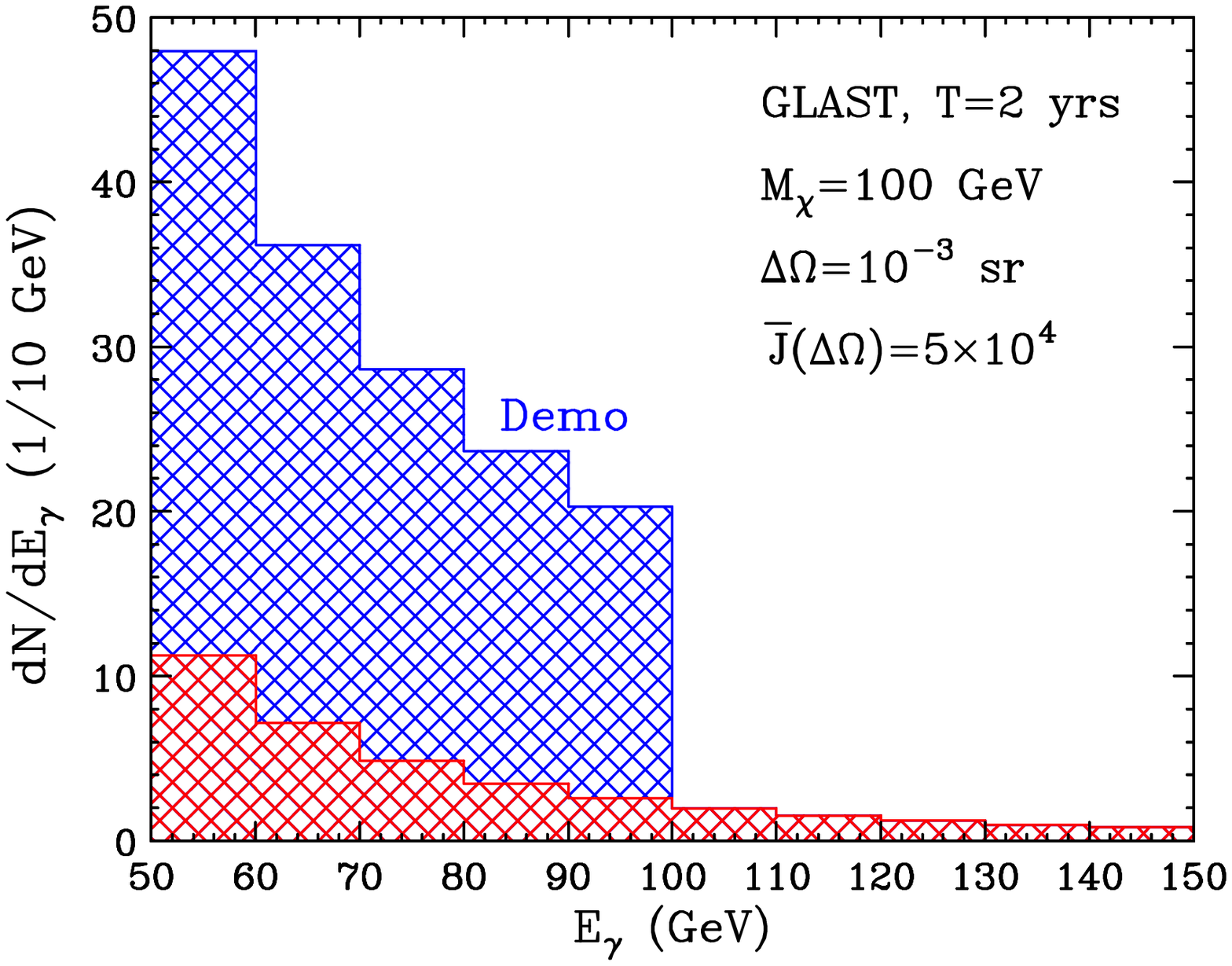}
\vskip2mm
\caption{The number of signal (blue) and background (red) events at the 
GLAST space telescope (collection area $A=10000$ cm$^2$,  exposure time 
$T=2$ years, field of view $\Delta\Omega=10^{-3}$). The signal is computed 
assuming the "democratic" scenario with a 100 GeV WIMP and a galactic model 
with $\bar{J}(\Delta\Omega)=5\times 10^{4}$.}
\end{center}
\label{fig:space}
\end{figure} 

An analogous plot illustrating the observability of the edge feature at the 
GLAST space telescope~\cite{GLAST} is shown in Fig.~6.  We have assumed a 
collection area $A=10000$ cm$^2$, an exposure time $T=2$ years, and a field 
of view\footnote{The field of view at GLAST can be varied between about 
$5\times 10^{-6}$ sr (the angular resolution of the telescope) and 2.3 sr 
(the full field of view). While larger values of $\Delta\Omega$ are 
advantageous from the point of view of statistics, focusing narrowly on the 
galactic center can lead to improved signal/background ratio if the dark 
matter density has a sharp peak at the center. However, reducing 
$\Delta\Omega$ substantially below 10$^{-3}$ typically results in insufficient 
statistics with the assumed collection area and exposure time.} 
$\Delta\Omega=10^{-3}$ sr. We have further assumed the ``democratic'' scenario 
with a 100 GeV WIMP, and a galactic model with 
$\bar{J}(\Delta\Omega)=5\times 10^{4}$. Again, the edge feature would be 
easily discernible for these parameters. 

In addition to the FSR photon flux plotted in Figs.~5 and~6, photons are also 
expected to be produced both by quark fragmentation and loop-induced 
$\chi\chi\to\gamma\gamma,\gamma Z$ annihilation processes. As we showed in 
Section~2, the fragmentation component is subdominant to the FSR flux near 
the endpoint, and therefore will not affect the edge feature. However, this 
component may dominate the flux at lower energies, in which case the edge 
feature would be accompanied by a sharp change in the slope of the spectrum. 
The monochromatic photon flux from $\chi\chi\to\gamma\gamma$ will contribute 
to the signal in the bin containing $E_\gamma=m_\chi$. This contribution is
also generally subdominant since $\sigma(\chi\chi\to\gamma\gamma)/\sigma
(\chi\chi\to X\bar{X}\gamma)\sim\alpha\sim10^{-2}$. If present, the line 
contribution will make the edge feature even sharper than our predictions 
based on the FSR flux alone.

\begin{figure}[tb]
\begin{center}
\includegraphics[width=9.5cm]{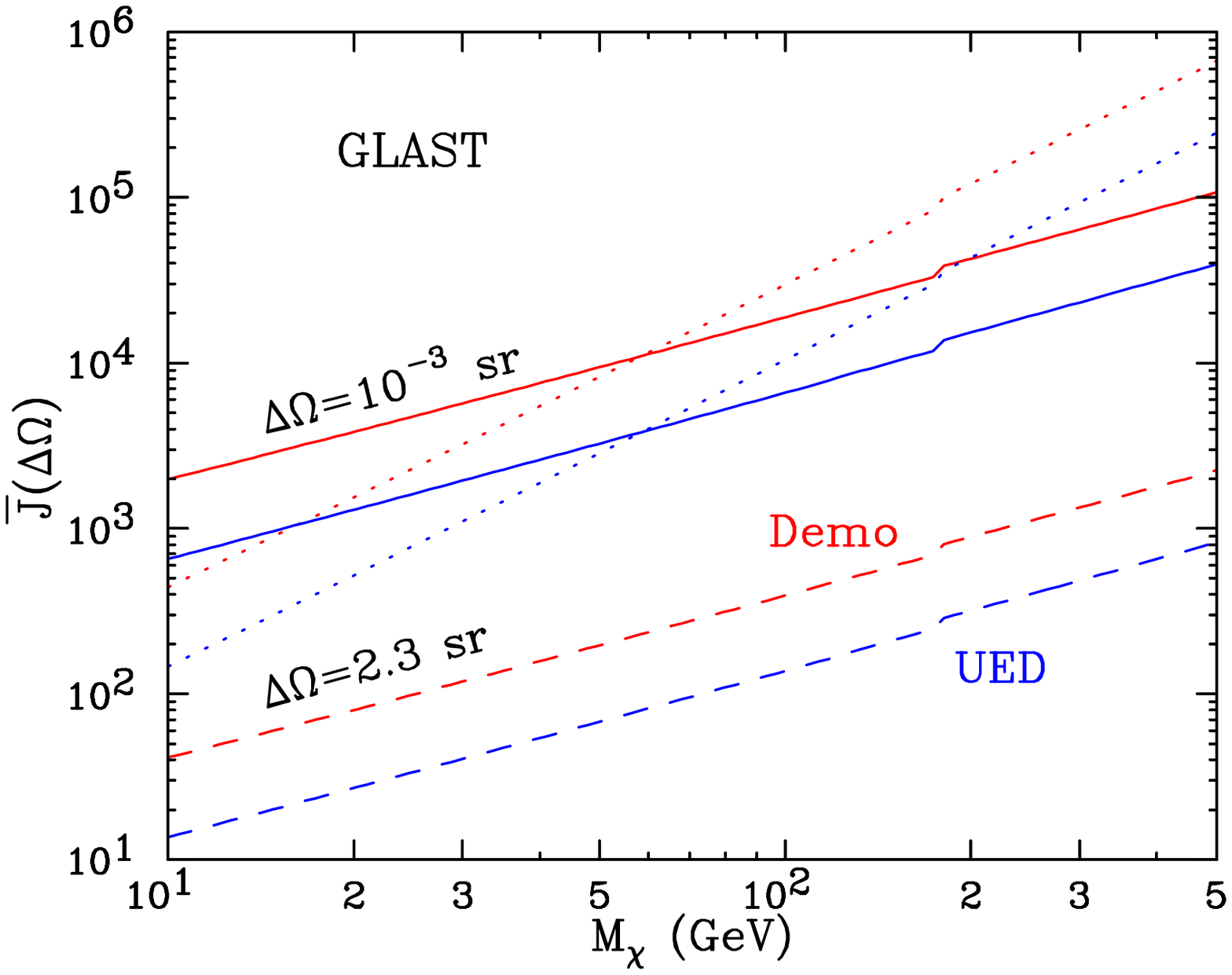}
\vskip2mm
\caption{The minimal value of $\bar{J}$ required for the discovery of the 
edge feature due to WIMP annihilation at the GLAST telescope (collection area 
$A=10000$ cm$^2$,  exposure time $T=2$ years, energy resolution 
$\delta=10$\%), for the field of view $\Delta\Omega=2.3$ sr (dash lines) and 
$\Delta\Omega=10^{-3}$ (solid lines show the minimum value of $\bar{J}$ for 
which a 3$\sigma$ deviation from the background occurs, while dotted lines 
represent the minimum value of $\bar{J}$ for which the edge bin contains at 
least 10 signal events.)}
\end{center}
\label{fig:GLASTreach}
\end{figure} 

To observe the FSR edge feature in the photon spectrum, the experiments 
need to search for a large drop in the number of events between two 
neighboring energy bins. A statistically significant discovery requires that 
the drop be larger than what can be expected from a fit to the rest of the 
spectrum. While a detailed analysis of the reach of any particular telescope 
is beyond the scope of this article, a simple estimate of the reach can be 
obtained as follows. Consider the energy bin  $[m_\chi(1-\delta), 
m_\chi(1+\delta)]$, where $\delta$ is the fractional energy resolution of a 
telescope\footnote{The assumption that  the bin is centered at $m_\chi$ 
represents the worst-case scenario for the reach; the reach can be improved by 
up to a factor of $\sqrt{2}$ by optimizing the binning to maximize
the significance. In addition, our estimates ignore the possible monochromatic 
photon flux from $\chi\chi\to\gamma\gamma$, which would appear in 
the same bin. The fragmentation photon flux, which is subdominant to the FSR 
component but could still enhance the signal, is also ignored. In this sense, 
our reach estimates are rather conservative.}. The number of signal events in 
this bin is
\beq
N_{\rm sig} \approx 1.4\times 10^{-12}\,g\,\delta\, \left(\frac{100~{\rm GeV}}
{m_\chi}\right)^2\bar{J}(\Delta\Omega) \,A_{{\rm cm}^2}
T_{\rm sec} \,\Delta\Omega\,,
\eeq{Nsig}
where $A_{{\rm cm}^2}$ and $T_{\rm sec}$ are the area of the telescope in 
cm$^2$ and the collection time in sec, respectively. 
Assuming that the fit to the high energy part of the spectrum ($E>m_\chi$)
produces an estimate of the background consistent with Eqs.~\leqn{sst_bgd} 
and~\leqn{act_bgd}, the expected number of background events $N_{\rm bg}$ in 
the energy bin $[m_\chi(1-\delta), m_\chi(1+\delta)]$ can be computed.
Requiring 
\beq
N_{\rm sig} \geq 3 \sqrt{N_{\rm bg}}
\eeq{stat}
for a statistically 
significant discovery of the step, we find that a discovery at a space-based
telescope is possible if
\beq
g \bar{J}(\Delta\Omega)\,\geq\,6\times 10^8 \,(A_{{\rm cm}^2} T_{\rm sec}
\delta \Delta\Omega)^{-1/2}\,\left(\frac{m_\chi}{100~{\rm GeV}}
\right)^{1.15}.
\eeq{spacereach}
This condition, together with the ``minimal signal'' requirement, 
\beq
N_{\rm sig}\geq 10, 
\eeq{minsig}
can be used to determine the reach of the GLAST 
telescope. The reach is shown in Fig.~7, where we 
plot the minimal value of $\bar{J}$ required for the discovery, as a 
function of the WIMP mass $m_\chi$, in the UED and ``democratic'' scenarios. 
The reach is shown for two values of $\Delta\Omega$: 2.3 sr, corresponding to 
utilizing the full field of view of the telescope, and 10$^{-3}$ sr, 
corresponding to focusing narrowly on the galactic center. (We assume the 
collection area $A=10000$ cm$^2$, the exposure time $T=2$ years, and 
the energy resolution $\delta=10$\%.) For $\Delta\Omega=2.3$ sr, the minimal 
signal criterion~\leqn{minsig} is always weaker than the 3$\sigma$ 
requirement in Eq.~\leqn{stat}, and we do not plot it. For 
$\Delta\Omega=10^{-3}$ sr, on the other hand, the minimal signal 
criterion~\leqn{minsig} dominates the reach determination for large masses; 
the dotted lines in Fig.~\ref{fig:GLASTreach} 
indicate the minimal value of $\bar{J}$ for which it is satisfied. Note that, 
while the reach in terms of $\bar{J}$ is clearly higher for the larger 
$\Delta\Omega$ due to higher statistics, the values of $\bar{J}$ in most 
galactic halo models are substantially enhanced at low values of 
$\Delta\Omega$. 

\begin{figure}[tb]
\begin{center}
\includegraphics[width=9.5cm]{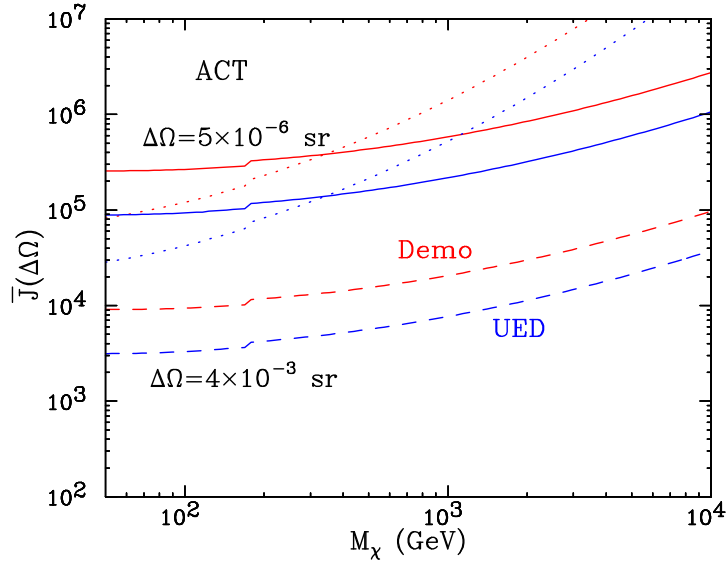}
\vskip2mm
\caption{The minimal value of $\bar{J}$ required for the discovery of the 
edge feature due to WIMP annihilation at a representative atmospheric 
Cerenkov telescope (collection area given in Eq.~\leqn{effarea}, exposure time 
$T=50$ hrs, energy resolution $\delta=15$\%), for the field of view 
$\Delta\Omega=4\times 10^{-3}$ sr (dash lines) and 
$\Delta\Omega=5\times10^{-6}$ sr (solid lines show the minimum value of 
$\bar{J}$ for which a 3$\sigma$ deviation from the background occurs, while 
dotted lines represent the minimum value of $\bar{J}$ for which the edge bin 
contains at least 10 signal events.)}
\end{center}
\label{fig:ACTreach}
\end{figure} 

The discovery reach for an ACT, assuming that the background is 
dominated by leptonic showers\footnote{The leptonic background is dominant 
over the entire range of WIMP masses of interest, provided that 
$\eps_{\rm had} \lapproxeq 0.1$.}, is given by
\beq
g \bar{J}(\Delta\Omega)\,\geq\,4\times 10^9 (A_{{\rm cm}^2} T_{\rm sec}
\delta \Delta\Omega)^{-1/2}\,\left(\frac{m_\chi}{100~{\rm GeV}}
\right)^{0.85}.  
\eeq{ACTreach}
To estimate the discovery potential of the VERITAS and HESS ACT arrays, 
consider an ACT with the collection area given in Eq.~\leqn{effarea}, an 
exposure 
time $T=50$ hrs, and the energy resolution $\delta=15$\%. The discovery reach 
for such a telescope is shown in Fig.~8. Dashed contours correspond to an 
experiment utilizing the full field of view of the ACT, assumed to be 
$4\times 10^{-3}$ sr. Solid contours indicate the reach of an experiment 
focusing narrowly on the galactic center, with the angular resolution of 
$0.07^\circ$ corresponding to $\Delta\Omega=5\times 10^{-6}$ sr. (Galactic 
halo model predictions for $\bar{J}$ for this value of $\Delta\Omega$ range 
from about $10^4$ to a few$\times10^8$.) In the latter case, the minimal 
signal requirement~\leqn{minsig} dominates the reach determination 
for large $m_\chi$, and is shown in the figure using the dotted lines. 

Given a model for galactic halo profile and a set of assumptions about the 
relevant annihilation fractions, Figs.~7 and~8 can be used to estimate the 
reach of the telescopes in terms of the highest value of the WIMP mass for 
which the edge feature can be observed. The estimate indicates that the 
prospects for observing the feature are quite good. For example, with a rather 
conservative assumption $\bar{J}(\Delta\Omega=4\times10^{-3}{~\rm sr})=10^4$, 
an ACT with the parameters used in our study would be able to discover the 
feature for $m_\chi$ up to about 2 TeV in the UED model, covering the entire 
range where the model is cosmologically consistent~\cite{KKDM}. Comparing 
Figs.~7 and 8 
indicates that the VERITAS and HESS arrays have a sensitivity comparable to 
GLAST. The experiments are complementary in terms of the range of WIMP masses 
that can be covered: the ACT will be sensitive to values of $m_\chi$ between 
about 50 GeV and 10 TeV, while GLAST can observe the FSR edge if $m_\chi$ is 
in the 10 -- 250 GeV mass range. We conclude that both space based telescopes 
and ACTs could provide sufficient sensitivity in the near future to discover 
the edge feature in the $\gamma$ flux if WIMPs are $s$-annihilators and the 
galactic halo profile and annihilation fractions are favorable.

Since the edge feature appears at $E_\gamma=m_\chi$, an observation of this 
feature would provide a direct measurement of the WIMP mass, with an accuracy 
determined by the energy resolution of the telescope, potentially better 
than 10\%. This is especially interesting because this parameter would be 
difficult to measure in a collider experiment, since WIMPs are pair-produced 
and escape the detector without interacting. Thus, observation of the edge 
feature would provide information complementary to what will be obtained at 
the LHC. For example, in the case of supersymmetry, the LHC can often 
determine the mass differences between some of the superpartners and the 
lightest neutralino, but not always the overall mass scale~\cite{Ian}. This 
ambiguity could be resolved if the edge feature in the gamma ray spectrum is 
observed.

%%%%%%%%%%%%%%%%%%%%%

\section{Conclusions} 
\label{conc}

In this article, we have obtained a prediction for the flux of photons
produced as final state radiation in galactic WIMP annihilation
processes. The prediction relies on the determination of the total WIMP
annihilation cross section, which is provided by the measurement of the
current cosmological dark matter abundance. As emphasized in~\cite{us},
this determination does not require any assumptions about the
fundamental physics giving rise to the WIMP, apart from the mild
condition of a generic mass spectrum. While the distribution of the
cross section among various possible final states is not constrained by
cosmological arguments, the FSR photons are produced for almost every
possible final state (with the exception of $\nu\bar{\nu}$ and $gg$),
making this signature quite model-independent. Moreover, if the final
state of WIMP annihilation is a pair of charged fermions (leptons or
quarks), the FSR flux has a well-defined step-like edge feature,
dropping abruptly at the energy equal to the WIMP mass. Observing such
a feature would provide strong evidence for the WIMP-related nature of
the flux distortion, and yield a measurement of the WIMP mass.
                                                                               
If WIMPs are $s$-annihilators, the predicted FSR fluxes can be quite
sizable, and the edge feature can be easily discernible above the
expected background. Using a rough statistical criterion, we have shown
that both ground-based ACTs such as HESS and VERITAS and space-based
gamma telescopes such as GLAST have a good chance of observing the edge
feature. It is likely that our simplified analysis underestimates the
ability of the experiments to observe a step-like feature in the photon
spectrum; a more sophisticated statistical analysis is clearly needed
to obtain a more realistic estimate of the reach.
                                                                              
In the $p$-annihilator WIMP case, the fluxes are expected to be lower,
and it is difficult to make model-independent predictions due to the
possible presence of an $a$ term in the annihilation cross section
which would not affect the WIMP relic abundance, but could dominate
galactic WIMP annihilation. Nevertheless, it would be interesting to
analyze if observable FSR photon fluxes can be produced in models with
$p$-annihilator WIMPs, such as the bino-like neutralino in supersymmetric 
models.
                                                                               
In summary, the flux of FSR photons emitted in the process of WIMP
annihilation in the center of Milky Way could be observable. An
observation of the step-like edge feature characteristic of this flux
could provide the first robust signature of WIMP dark matter.
We encourage the collaborations involved in the analysis of the data
coming from ground-based and space-based gamma ray telescopes to
perform systematic searches for this important signature in a model-independent
fashion as presented here.

%%%%%%%%%%%%%%%%%%%%%

\section*{Acknowledgments} 

We are grateful to the Aspen Center for Physics where part of this work has been completed. MP and AS are supported by the
NSF grant PHY-0355005. KM and AB are supported by a US DoE 
OJI award under grant DE-FG02-97ER41029.

\end{document}